\documentclass[a4paper,fleqn,usenatbib]{mnras}
\usepackage[T1]{fontenc}
\usepackage{ae,aecompl}
\usepackage{amssymb}
\usepackage{amsmath}
\usepackage{mathptmx}
\usepackage{epstopdf}
\usepackage{booktabs}
\usepackage{float}
\usepackage{subfig}
\usepackage{epsfig,color,ulem}
\usepackage{tabularx}

\newcommand{\A}{{\it A}}
\newcommand{\B}{{\it B}}
\newcommand{\C}{{\it C}}
\newcommand{\D}{{\it D}}
\newcommand{\E}{{\it E}}
\newcommand{\F}{{\it F}}
\newcommand{\G}{{\it G}}
\newcommand{\h}{{\it H}}

\newcommand{\msun}{\,{\rm M_{\odot}}}
\newcommand{\cm}{\,{\rm cm}}

\newcommand{\erg}{\,{\rm erg}}

\newcommand{\lmax}{\,{L_{\rm max}}}
\newcommand{\lmin}{\,{L_{\rm min}}}

\newcommand{\s}{\,{\rm s}}
\def\gsim{ \lower .75ex \hbox{$\sim$} \llap{\raise .27ex \hbox{$>$}} }
\def\lsim{ \lower .75ex\hbox{$\sim$} \llap{\raise .27ex \hbox{$<$}} }
\def\app#1#2{%
	\mathrel{%
		\setbox0=\hbox{$#1\sim$}%
		\setbox2=\hbox{%
			\rlap{\hbox{$#1\propto$}}%
			\lower1.1\ht0\box0%
		}%
		\raise0.25\ht2\box2%
	}%
}

\title[No GRBs from intermittent hydrodynamic jets]{Intermittent hydrodynamic jets in collapsars do not produce GRBs}
\author[Gottlieb, Levinson \& Nakar]{
	Ore Gottlieb\thanks{oregottlieb@mail.tau.ac.il},
	Amir Levinson,
	Ehud Nakar
	\\
	{School of Physics and
		Astronomy, Tel Aviv University, Tel Aviv 69978, Israel}
}
\pubyear{2020}
\begin{document}
	\label{firstpage}
	\pagerange{\pageref{firstpage}--\pageref{lastpage}}
	\maketitle	
	\begin{abstract}
		Strong variability is a common characteristic of the prompt emission of gamma-ray bursts (GRB).	This observed variability is widely attributed to an intermittency of the central engine, through formation of strong internal shocks in the GRB-emitting jet expelled by the engine. In this paper we study numerically the propagation of hydrodynamic jets, injected periodically by a variable engine, through the envelope of a collapsed star. By post-processing the output of 3D numerical simulations, we compute the net radiative efficiency of the outflow. We find that all intermittent jets are subject to heavy baryon contamination that inhibits the emission at and above the photosphere well below detection limits.
		This is in contrast to continuous jets that, as shown recently, produce a highly variable gamma-ray photospheric emission with high efficiency, owing to the interaction of the jet with the stellar envelope. Our results challenge the variable engine model for hydrodynamic jets, and may impose constraints on the duty cycle of GRB engines. If such systems exist in nature, they are not expected to produce bright gamma-ray emission, but should appear as X-ray, optical and radio transients that resemble a delayed GRB afterglow signal.
	\end{abstract}
	
	\section{Introduction}
	\label{sec:introduction}	
	
	The prompt GRB emission exhibit variability over a vast range of timescales - from milliseconds to seconds \citep{McBreen1994,Li1996,Norris1996,Ramirez-Ruiz1998,RamirezRuiz2000,Nakar2002b,Nakar2002c}, with a large scatter of characteristics across sources. This observed variability may reflect an intermittent activity of the central engine  \citep[e.g.][]{Levinson1993,Sari1997b,MacFadyen1998,Fenimore1999,Aloy2000,Lin2016}, may result from the interaction between the jet and the medium into which it is launched \citep[e.g.][hereafter GLN19]{Aloy2002,Matzner2003,Morsony2007,Gottlieb2019b}, or both.
	
	Recently, we have shown (GLN19) that under conditions anticipated in most GRBs, continuous injection of a hydrodynamic jet leads to an efficient photospheric emission, owing to a strong dissipation of the flow at a collimation shock which is located at a large distance from the jet injection point (roughly the stellar radius in long GRBs or the edge of the merger ejecta in short GRBs). 
	The  analysis of GLN19 further confirms that the photospheric efficiency of individual fluid elements is very sensitive to their baryon load; relatively low-loaded elements with terminal Lorentz factor $\Gamma \gtrsim 100$ show high photospheric efficiencies, while more heavily loaded elements with lower Lorentz factors show low efficiencies.
	The 3D numerical simulations performed in GLN19 also indicate that the jet-medium interaction induces a rapid onset of the Rayleigh-Taylor instability, as previously found \citep{Meliani2010,Matsumoto2013a,Matsumoto2013,Matsumoto2019,Matsumoto2017,Toma2017,Gourgouliatos2018}, which in turn gives rise to a strong mixing of jet and cocoon material.
	This sporadic mixing results in large variations in the baryon loading of the different fluid elements, which in turn leads to large variations in the radiative efficiency that is seen as rapid variability of the photospheric emission.
	While the mixing enhances the mean baryon load on jet streamlines, the Lorentz factor within the jet core (a few degrees) was found to be still large enough ($\Gamma>100$) in all cases explored to allow a high average efficiency (albeit with large fluctuations) of the photospheric emission.  However, energy dissipation near the photoshpere by internal shocks that form
	in the mixing process has been found to be rather small, and the question whether it can significantly modify the emergent spectrum remains open. 

	Considerable dissipation near or above the photosphere in a weakly magnetized outflow seems to require the formation of strong internal shocks by intermittency of the central engine (in difference from reconnection in a Poynting jet). 
	Such intermittency is also naively anticipated on physical grounds, as it is difficult to envision a steady operation of the engine over times vastly longer than the dynamical time.  Since the outflow must break out of the confining medium prior to emitting, the question arises as to how the modulations produced by the engine evolve as they propagate through the surrounding medium.  In this paper we address this question by performing 3D numerical simulations of GRB jets with periodic injection.

	Previous works \citep{Morsony2010,LopezCamara2014,Geng2016,Parsotan2018} attempted to compute the structure and emission of a modulated outflow by performing 2D simulations.
	However, axisymmetric simulations are known to be prone to severe numerical artifacts and, therefore, results based on such simulations cannot be trusted.  In particular, the aforementioned mixing is absent in 2D simulations by virtue of the complete suppression of non-axisymetric instabilities (see e.g. discussion in GLN19). 
	Full 3D simulations of intermittent jet launching were performed recently by \citet{LopezCamara2016}. They found that shorter periodicity of the engine leads to heavier baryon loading and, consequently, lower terminal 
	Lorentz factor.  However, their simulation box extends only up to several stellar radii, much below the photosphere, and
	it is unclear from their results how the loading affects the prompt emission.  
	%and the interaction of the jet with the medium shortens the duration of bright episodes. Their results also feature a moderate Lorentz factor of the flow due to the heavy loading. However, in their study they have not considered the resulting prompt emission from the mixed outflow.}
	
	Here we perform 3D RHD simulations of modulated outflows, with modulation periods much shorter than the breakout time from the star, and follow their propagation from the injection radius, well
	within the star, up to several stellar radii.  The output of the 3D simulations is then used to propagate the flow, using 2D simulations, to the vicinity of the photosphere.  We find that such 
	short modulations
	%to calculate the emission from the widely popular variable jet's engine model. We consider a variety of duty cycle timescales that are much shorter than the breakout time of a lGRB jet from a collapsing massive star.
	%We show that the modulations in the jet 
	induce, quite generally, a substantially larger mixing of jet and stellar material along the jet axis 
	than in the case of a continuous jet.  This overloading of the jet pushes the photosphere to radii much 
	larger than the dissipation radius of internal shocks generated by collisions of fast and slow shells,
	leading to extremely small (practically zero) radiative efficiency. We thus conclude that intermittent engines that 
	expel hydrodynamic flows cannot produce the observed GRBs.  If such systems exist in nature they should
	appear as X-ray, optical and radio transients that resemble GRB afterglows starting an hour or so after the burst.
	%	However, the emerging structure gives rise to a delayed on-axis afterglow.
	
	\section{Numerical Setup}
	\label{sec:setup}
	
	We perform 3D RHD simulations of long GRB jets with an intermittent injection (see table \ref{table}).
	In all shown simulations a top-hat jet with an initial opening angle $ \theta_0 = 8^\circ $ and a cylindrical radius $ r_0 = 10^8\cm $ is injected from $ z_{\rm{beg}} = r_0/\theta_0 = 7\times 10^8\cm $. 
	The Lorentz factor and enthalpy per baryon at the injection boundary are, respectively, $ \Gamma_0 = 5 $ and $ h_0 = 100 $ at all times, such that the terminal Lorentz factor is limited to $ \eta_0 \equiv h_0\Gamma_0 = 500 $.  
	The jet power is varied by changing the density at the injection boundary.  To be concrete, the power is given 
	by $L_j(t) = \rho_{j0}(t) h_0\Gamma_0^2\beta_0 (\pi r_0^2)c^3$, where the injected density $\rho_{j0}(t)$ is a periodic square wave in models $\A-\C$ and $\E-\G$, with period and amplitude as given in table \ref{table}, and is a square of a sinusoidal function in model $D$.  Our reference model $\h$ corresponds to a continuous jet injection with a power equals to the mean of that of models $E$ and $G$. 
	%The jet maintains a total two-sided luminosity $ \lmax = 10^{50}\erg\s^{-1} $, initial Lorentz factor $ \Gamma_0 = 5 $ and initial specific enthalpy $ h_0 = 100 $ such that the terminal Lorentz factor is $ \Gamma_\infty \equiv h_0\Gamma_0 = 500 $. 
	The jet is launched	into a static non-rotating star with a mass of $ M = 10\msun $, radius of $ R_\star = 10^{11}\cm $ and a density profile $ \rho(r) \propto r^{-2}x^3 $, where $ x \equiv (R_\star-r)/R_\star $.
	%In our simulations we change the luminosity periodically, either by the square of a sinusoidal function or by a Heaviside step-function. We examine the effect of two parameters: the time period of the duty-cycle, $ T $, and the minimal luminosity in the cycle, $ \lmin $.
	%As a reference model we also consider a continuous injection of a jet with the same total energy as that of the modulated jets.
	%Additionally, we consider one intermittent sGRB model which is based on simulation $ \A $ in \citet{Mooley2018b}.
	%The full setups are listed in Table \ref{table}.
	We remind the reader that the product $\eta=h\Gamma$ is conserved along streamlines of an adiabatic hydrodynamic flow (including across shock fronts). Consequently, in the absence of baryon loading and/or radiative losses it must remain constant.  In our analysis, any reduction in $\eta$ along streamlines indicates local mass entrainment. 
	Thus, in what follows we use $\eta$ as a measure for mixing (see GLN19 for further details).  
	
	Our 3D simulation grids end at $ 5R_\star $. Since fresh elements show less baryon loading over time (GLN19), we continue the simulations after the jet head reaches $ 5R_\star $ to study the temporal evolution of the mixing in new jet elements.
	For models $ \E $ and $ \G $, which hold observational promise, we trace the elements further, up to $ 100R_\star $. For that purpose, we convert our 3D grids to 2D once the jet head reaches the edge of the 3D simulation box \citep[see method in][]{Gottlieb2018b}. At this point we stop the injection of the jet so that the jet length is $ 5R_\star $.
	We stress that above $ z = 5R_\star $ 2D artifacts are not expected to have a substantial effect on the jet evolution \citep{Gottlieb2018b}.
	
	The 3D grids are Cartesian and identical to each other. We use three patches along the $ x $ and $ y $ axes, and one patch on the $ z $-axis along which the jet propagates. The inner $ x $ and $ y $ axes are in the inner $ |7.5\times 10^8\cm| $ with 60 uniform cells. The outer patches are stretched logarithmically to $ |1\times 10^{11}\cm| $ with 120 cells on each side. The $ z $-axis has one uniform patch from $ z_{\rm{beg}} $ to $ 5R_\star $ with 4000 cells. The total number of cells is therefore $ 300\times 300\times 4000 $.
	The 2D grids are cylindrical with the $ r $ axis having the same resolution as the $ x $ and $ y $ axes in the 3D. The $ z $ axis is extended to $ 10^{13}\cm $ while keeping the same cell resolution, i.e. 80,000 uniform cells from $ z_{\rm{beg}} $ to $ 100R_\star $. The total number of cells is therefore $ 300\times 80000 $. Convergence tests are shown in Appendix \ref{sec:convergence}.
	%For the sGRB grid setup, we use three patches for the $ x $ and $ y $ axes, which includes an inner uniform grid of 80 cells inside $ | 10^8\cm | $ and an outer logarithmic grid from  $ | 10^8\cm | $ to $ 10^{10}\cm $ with 300 cells. The $ z $-axis is uniform with 3000 cells from $ 4.5\times 10^8\cm $ to $ 2\times 10^{10}\cm $. In total the sGRB simulation includes $ 680\times680\times 3000 $ cells.
	
	\begin{table}
		\setlength{\tabcolsep}{6.2pt}
		
		\centering
		\begin{tabular}{ | l | c  c  c  c | }
			
			\hline
			Model & $ f(t) $ & $ T [\s] $ & $ \lmax [10^{50}\erg] $ & $ \lmin [10^{50}\erg] $ \\ \hline
			$ \A $ & $ \Theta(t) $ & $ 0.2 $ & 1.0 & $ 0 $ \\
			$ \B $ & $ \Theta(t) $ & $ 2.0 $ & 1.0 & $ 0 $\\
			$ \C $ & $ \Theta(t) $ & $ 4.0 $ & 1.0 & $ 0 $ \\
			$ \D $ & $ {\rm sin}^2(t) $ & $ 2.0 $ & 1.0 & $ 0 $ \\
			$ \E $ & $ \Theta(t) $ & $ 0.2 $ & 1.0 & $ 0.2 $ \\
			$ \F $ & $ \Theta(t) $ & $ 2.0 $ & 1.0 & $ 0.1 $ \\
			$ \G $ & $ \Theta(t) $ & $ 2.0 $ & 1.0 & $ 0.2 $ \\
			$ \h $ & $ \Theta(t) $ & $ \infty $ & 0.6 & $ 0.6 $ \\
			\hline
		\end{tabular}
		\hfill\break
		
		\caption{The configurations of the simulations. $ f(t) $ is the injected periodic function: Heaviside step-function $ \Theta(t) $ or $ {\rm sin}^2(t) $, $ T $ is the duty cycle time, and $ \lmax (\lmin) $ are the maximal (minimal) injected luminosity}\label{table}
	\end{table}
	
	\section{Hydrodynamics}
	\label{sec:hydro}
	
	\begin{figure*}
		\centering
		\includegraphics[scale=0.4]{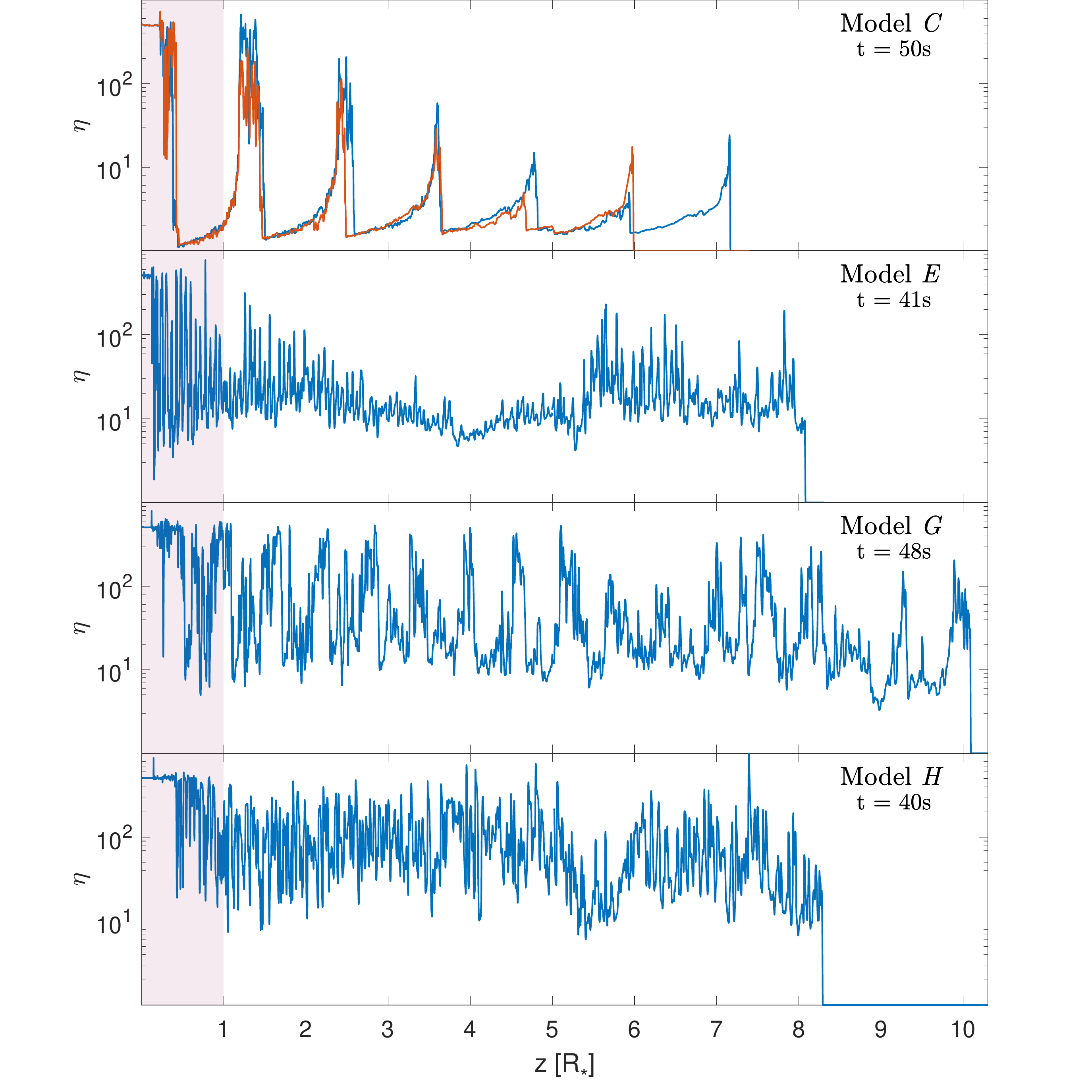}
		\caption[]{
			Profiles of $\eta$ (the terminal Lorentz factor of each fluid element if it were propagating undisturbed to infinity) along the jet axis at the end of the 3D simulation for a sample of models. The indicated times are measured in the rest frame of the star.  The pale red stripe marks the interior of the star.
			The red line in the top panel delineates the profile one duty cycle (4s) earlier than the blue line, indicating no evolution in time. Since our grid is stretched up to $ 5R_\star $, the data at $ z > 5R_\star $ is shifted from earlier times, assuming that no further mixing occurred.}
		\label{fig:hg_profiles}
	\end{figure*}
	
	All simulations were run for times much longer than the breakout time, up to a point 
	at which no significant change in the evolution of newly ejected shells is seen in each case.
	In all models we find that intermittent jet injection leads to excessive loading of the jet by mixing relative
	to the reference model, $\h$, where the jet is injected  continuously.  This is seen in 
	Figure \ref{fig:hg_profiles}, where $\eta$ profiles along the jet axis are plotted at the 
	end of the simulation time (as indicated in each panel)\footnote{Since the 3D simulation box extends only 
		up to $5R_\star$, by the end of the simulation some of the fluid at the outflow front has propagated outside the box. The extension beyond $5R_\star$ shown in Fig \ref{fig:hg_profiles} was obtained by extrapolating the 3D solution at earlier times to these radii.}. 
	
\begin{figure}
		\centering
		%	\begin{frame}{}
		
		%		\includemedia[width=1\linewidth,height=1\linewidth,activate=pageopen,
		%passcontext,
		%transparent,
		%addresource=Modulations01.mp4,
		%flashvars={source=Modulations01.mp4}
		%]{\includegraphics[width=0.6\linewidth]{3dmaps.pdf}}{VPlayer.swf}
		%		flashvars={
		%			modestbranding=1 % no YT logo in control bar
		%			&autohide=1 % controlbar autohide
		%			&showinfo=0 % no title and other info before start
		%			&rel=0 % no related videos after end
		%		}
		%		]{\includegraphics[width=0.6\linewidth]{3dmaps.pdf}}{https://youtu.be/UlEJ2MYDlcE}
		
		%	\end{frame}
		\includegraphics[width=1\linewidth,height=1\linewidth]{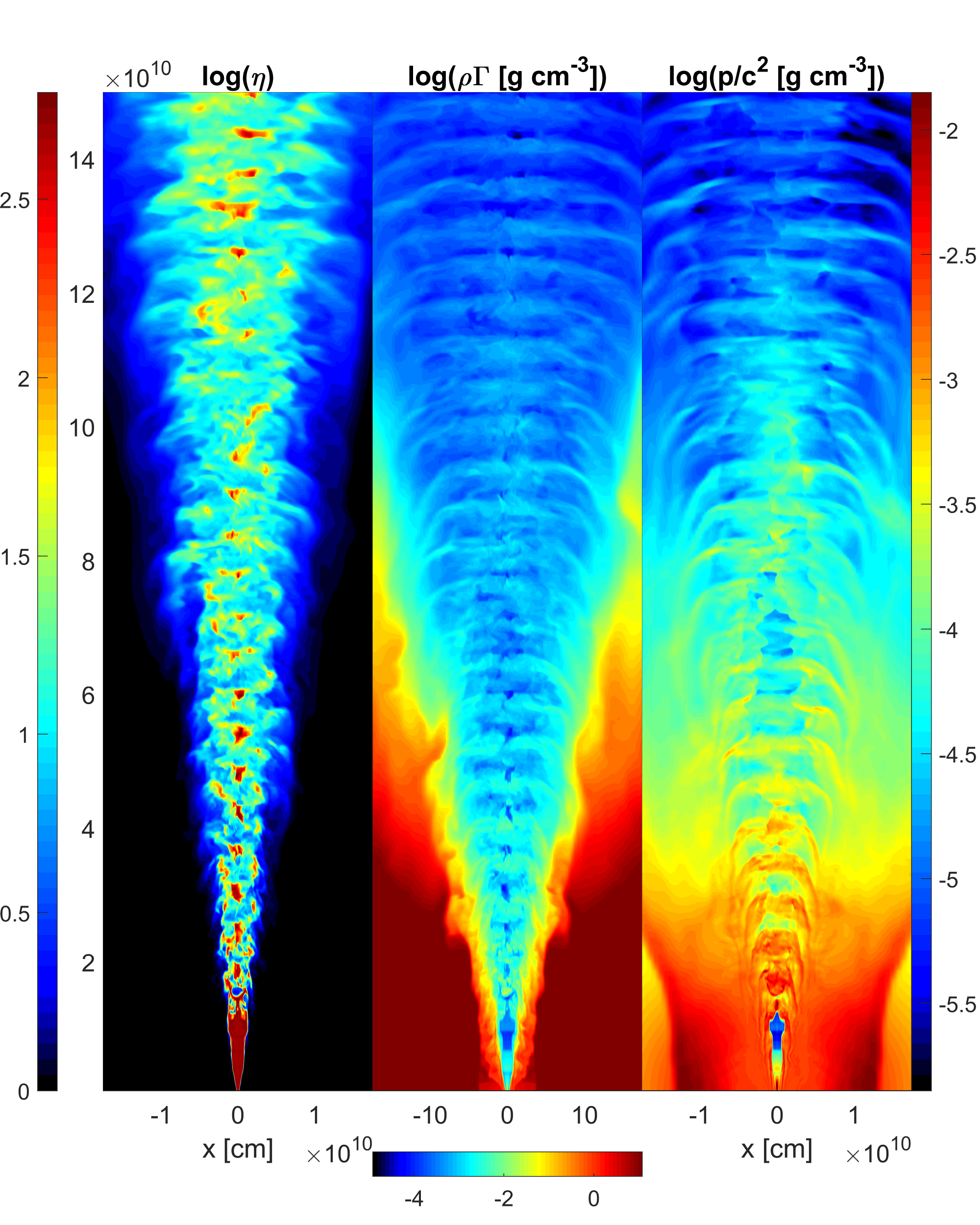}
		\includegraphics[width=1\linewidth,height=1\linewidth]{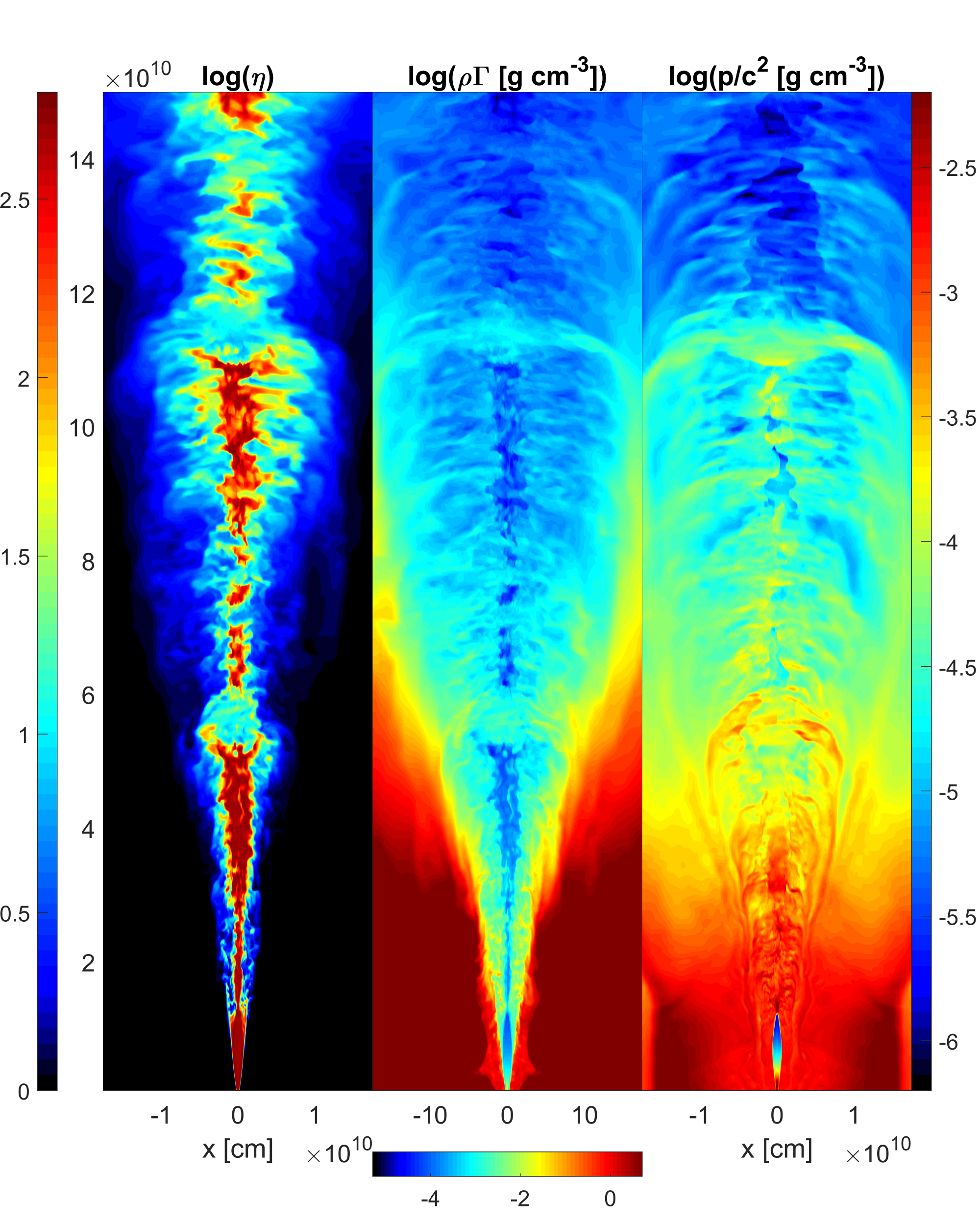}
		\caption[]{
			Maps of model $ \E $ (top) and $ \G $ (bottom) on plane $ x-z $ up to $ z = 1.5R_\star $ at times 41s and 48s, respectively. Shown are $\eta$ (left), mass density in the star frame (middle) and pressure (right). A video for model $ \E $ is available at \url{http://www.astro.tau.ac.il/~ore/instabilities.html\#modulations}.
		}
		\label{fig:3dmaps}
	\end{figure}

	The excessive mixing in intermittent jets seems to be a result of a different mixing process than the one seen in a continuous jet, as we explain below. When a continuous jet propagates in the dense medium (before it breaks out), a forward-reverse shock structure forms at its head. Any jet material that enters this structure via the reverse shock is heavily mixed. After the jet breaks out, the head material is pushed sideways and the jet is free to propagate uninterrupted. The mixing is then dominated by the R-T instability that develops above the collimation throat along the jet-cocoon interface. In a periodic jet the picture is different. The mixing of jet and stellar material is dominated by entrainment of ambient matter through the shock that forms at the head of each high-power shell,(see Figure \ref{fig:3dmaps} for
	models $E$ and $G$), rather than by the formation of R-T 
	instabilities at the collimation throat.
	
	In models where each cycle of jet ejection is followed by a quiescent phase ($A-C$), we find that during each quiescent episode the cavity opened by the jet is filled by stellar material fast enough before the next cycle of jet material arrives. Consequently, a repeated forward-reverse shock structures (similar to the head in the continuous jet case) is formed at the front of each shell ejected in a new cycle, where the relativistic jet pushes upon the heavy stellar material. The mixing at these ``heads" is high. Since the high-power shells do not have sufficient energy to evacuate the large mass they encounter, they are choked in the dense medium early on, powering a non-relativistic expanding shock in the star.
	The mass enclosed inside the cavities between shells is dragged along with the outflow long after breakout occurs, as seen in the top panel of Figure \ref{fig:hg_profiles}  (low values of $\eta$ indicate stellar material whereas high values correspond to the high-power shells). Ultimately, the fast shells decelerate, the slow dense shells accelerate, and the entire outflow approaches a roughly uniform Lorentz factor. We find the mean Lorentz factor on the jet axis at the end of the simulation to be $ <\eta> \equiv \int{dE}/\int{dM} \lesssim 10 $ for models $A-C$, 
	compared with $ <\eta> \approx 90 $ for the reference model $H$.  The heavy baryon load of the intermittent jet
	renders the radiative efficiency at the photosphere to be practically zero in these models.

	%The remaining models show qualitatively similar behaviour.  Here, there is no empty cavity that can be quickly
	%filled with ambient matter, as in models $A-C$.  Nonetheless, the low-power jet sections are over-compressed 
	%by the cocoons pressure, resulting in the formation of a shock at the head of each high-power shell that 
	%drags external mass.  

	%The aforementioned evolution results in a non-relativistic shock breakout from the star without any relativistic jet material \citep[see also][]{LopezCamara2016}. Consequently, the dense medium at which the jet propagates is no longer embedded in the star, but expands farther. Namely, even after jet elements break out from the star, they still interact with the dense expanding material, so that they continue to mix and lose energy after breaking out as well.
	%This behavior can be seen in the top panel of Figure \ref{fig:hg_profiles} which depicts the value of the terminal Lorentz factor $ \Gamma_\infty $ along the jet axis for model $ \C $, assuming no further mixing takes place. The blue and red lines depict successive duty cycles, i.e. at times that differ by four seconds. One can see that the injected jet material undergoes intense mixing, both inside and outside the star, and consequently its $ \Gamma_\infty $ decreases asymptotically to zero. Furthermore, no improvement in maintaining a high value of $ \Gamma_\infty $ is seen over time.
	
	%Our simulations show that longer duty cycles lead to less baryon loading and higher Lorentz factors in the jets, similar to what was found by \citet{LopezCamara2016}. One possible explanation is that during every quiescent episode, the baryon entrainment has enough time to get through the jet, even for the shortest duty cycles. In such case longer duty cycles have less quiescent periods and hence remain more stable.
	%It is possible however that for a sufficiently short duty cycle, there will not be enough time for baryon entrainment into the jet, so that the jet will evolve in a similar way to a continuous jet.
	%That implies that there is a certain duty cycle time period for which a maximal entrainment of baryons is obtained.
	%Unfortunately, modeling duty cycles of $ T \lesssim 10^{-2}\s $ requires simulations with a very high resolution, which is currently too demanding for our available computational power.
	
	%The heavy loading decelerates all jets with $ \lmin = 0 $ (models $ \A, \B, \C $) to energy-weighted average asymptotic Lorentz factor of $ \bar{\Gamma}_\infty \approx 10 $. This result is much below current estimates of GRB jets, and thus such jets cannot produce any detectable $ \gamma $-ray signals.
	
	In models $ \E $, $ \F $ and $ \G $ (model $ \D $ is found to be similar to model $ \G $), where the jet power never falls below $ 10\% $ of its maximal value, there are no quiescent phases during which stellar material can fill up the cavity in the jet. However, the modulation of the jet power leads to over-compression of low-power jet sections by the pressure in the cocoon, resulting, similarly to the cases where the jet has quiescent phases, in the formation of a forward-reverse shock at the head of each 
	high-power shell that once again leads to mass entrainment (see Figure \ref{fig:3dmaps}).   For a given modulation amplitude, we generally find larger mixing at shorter modulation periods. 
	%The effect of $ \lmin $ on $ \Gamma_\infty $ profile on the axis is demonstrated in Figure \ref{fig:hg_profiles}. Models $ \E $ and $ \G $, in which $ \lmin = 0.2\lmax $, feature a variable profile which maintains $ \Gamma_\infty \gtrsim 10 $. 
	The mean asymptotic Lorentz factor is $ <\eta> \approx $ 15, 20 and 25 for simulations $ \E $, $ \F $ and $ \G $, respectively; higher than in cases $A, B, C$ but still significantly lower than that of the reference continuous jet in model.  We further find that internal shocks which are generated by collisions of slow and fast shells decay well below the photosphere, so that significant dissipation at or above the photosphere does not ensue in these models. 
	
	Figures \ref{fig:hg_profiles} and \ref{fig:3dmaps} also show fast variations in $\eta$ superposed on the periodic modulations in models $E$ and $G$. These are caused by the instability of the collimating flow, as in the continuous jet model $H$.  However, as stated above, loading by this process is sub-dominant.   In principle, these variations  
	can be imprinted in the resulting photospheric emission, however, as shown in the next section, the radiative efficiency in all modulated jet models is too small for the prompt emission to be detected.
	
\begin{figure}
		\centering
		\includegraphics[scale=0.21]{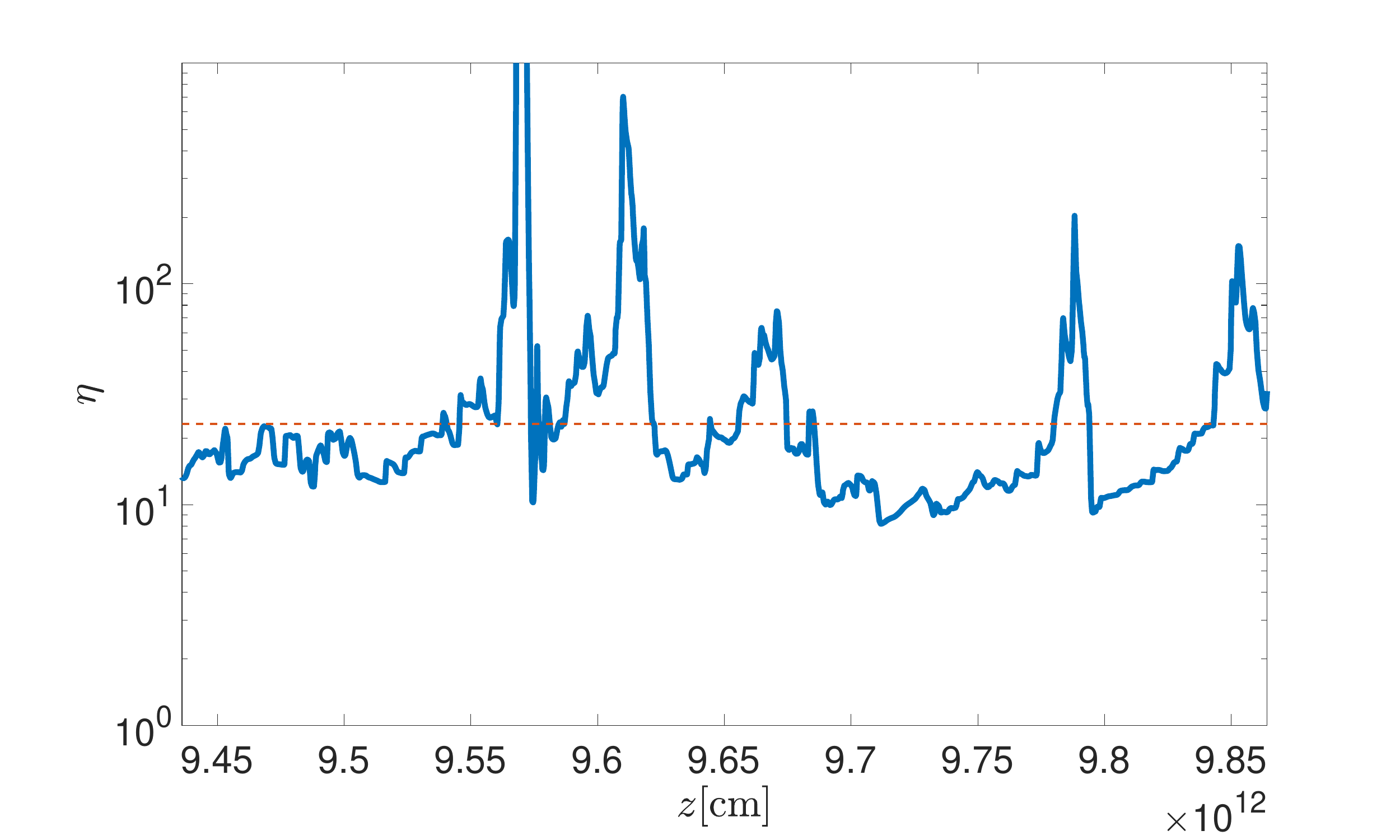}
		\caption[]{
			The $ \eta $ profile on the jet axis at the last snapshot of the 2D simulation of model $ \G $, when the jet head reaches $ z \approx 10^{13}\cm $. The back and the front of the jet are cut as they are subject to numerical artifacts. At $ z = 9.57\times 10^{12}\cm $ a small slab of matter shows a nonphysical behavior with $ \eta \approx 10^5 $. This slab of matter has a negligible amount of energy and thus does not affect the hydrodynamics.
			The dashed red line depicts the value of $ <\eta> \approx 25 $ of the shown jet elements.
		}
		\label{fig:2d_profile}
	\end{figure}

	In Figure \ref{fig:hg_profiles} the fourth panel shows that at several stellar radii the flow in model $ \G $ is rather inhomogeneous due to the intermittent nature of the jet engine, and that a non-negligible fraction of the outflow has high Lorentz factor ($\gtrsim 100$). Over time, however, the internal shocks between fast and slow elements average out the hydrodynamic properties of the system, so that the flow becomes more homogeneous, approaching $<\eta>$ everywhere. In periodic jets, due to the low value of $<\eta>$, these internal shocks take place far below the photosphere and therefore have a small contribution to the photospheric gamma-ray efficiency. Figure \ref{fig:2d_profile} depicts the $ \eta $ profile on the jet axis in model $ \G $ when the jet head reaches $ z \approx 10^{13}\cm $. By this time the fast shocks have decelerated due to their interaction with the slow heavy elements. Consequently, the flow shown in the figure shows less variations between elements. At later times the flow will become even smoother by converging to the value of $ < \eta > $ (dashed red line). We note that in the 2D simulation a numerical artifact leads to the formation of a slab with $ \eta \approx 10^5 $, as shown in the figure. This slab contains a negligible amount of energy and thus does not affect the hydrodynamics, and can be ignored in the radiation calculation in the next section.
	
	\section{Emission}
	\label{sec:emission}
	
	\subsection{Gamma rays}
	\label{sec:gamma}
	
	We calculate the radiative efficiency at the photosphere for models $ \E $ and $ \G $ for which we
	anticipate it to be the highest among the modulated jet models in table \ref{table}, since
	$\eta$ has the largest values for these two models.
	The method employed is described in detail in GLN19. In short, 
	in GLN19 we applied the fireball equations on the last snapshot of the simulation to extrapolate the evolution of each fluid element along radial streamlines up to its photosphere $r_{ph}$, defined as the location at which the radial optical depth to infinity is unity, that is, $ \tau(r_{ph})=1 $. The radiative efficiency of the fluid element is estimated to be the value of the quantity $ q = (h-1)/h $ at $r_{ph}$, specifically $\epsilon=(h_{ph}-1)/h_{ph}$, 
	where $ h $ is the enthalpy per baryon and $h_{ph}=h(r_{ph})$.
	This extrapolation ignores possible dissipation in sub-photospheric internal shocks, which was justified for the 
	weak shocks generated by mixing in the continuous jet considered in GLN19 (model $H$ in table \ref{table}). 
	However, in models $\E$ and $\G$ the modulation of the engine gives rise to much stronger shocks. To properly 
	account for the evolution of the flow in these cases we perform 2D simulations in a box that extends up to $100R_\star$,
	where internal shocks decay, using the output of the 3D simulations as initial data for 2D simulations \citep[see method in][]{Gottlieb2018b}. Only above this radius we apply the extrapolation used in GLN19 to the photosphere.  We note that 
	the evolution of the outflow in the 2D simulations is followed for times much longer than the engine operation time, which in these runs is $\sim 30$s (the time it takes the jet to reach the outer edge of the 3D simulation box).  As noted in
	the preceding section, we find that the evolution of newly ejected shells does not change much after this time, implying that the properties of the outflow computed below, and in particular the radiative efficiency, are representative also for prolonged engine activity. 
	
\begin{figure}
		\centering
		\includegraphics[scale=0.26]{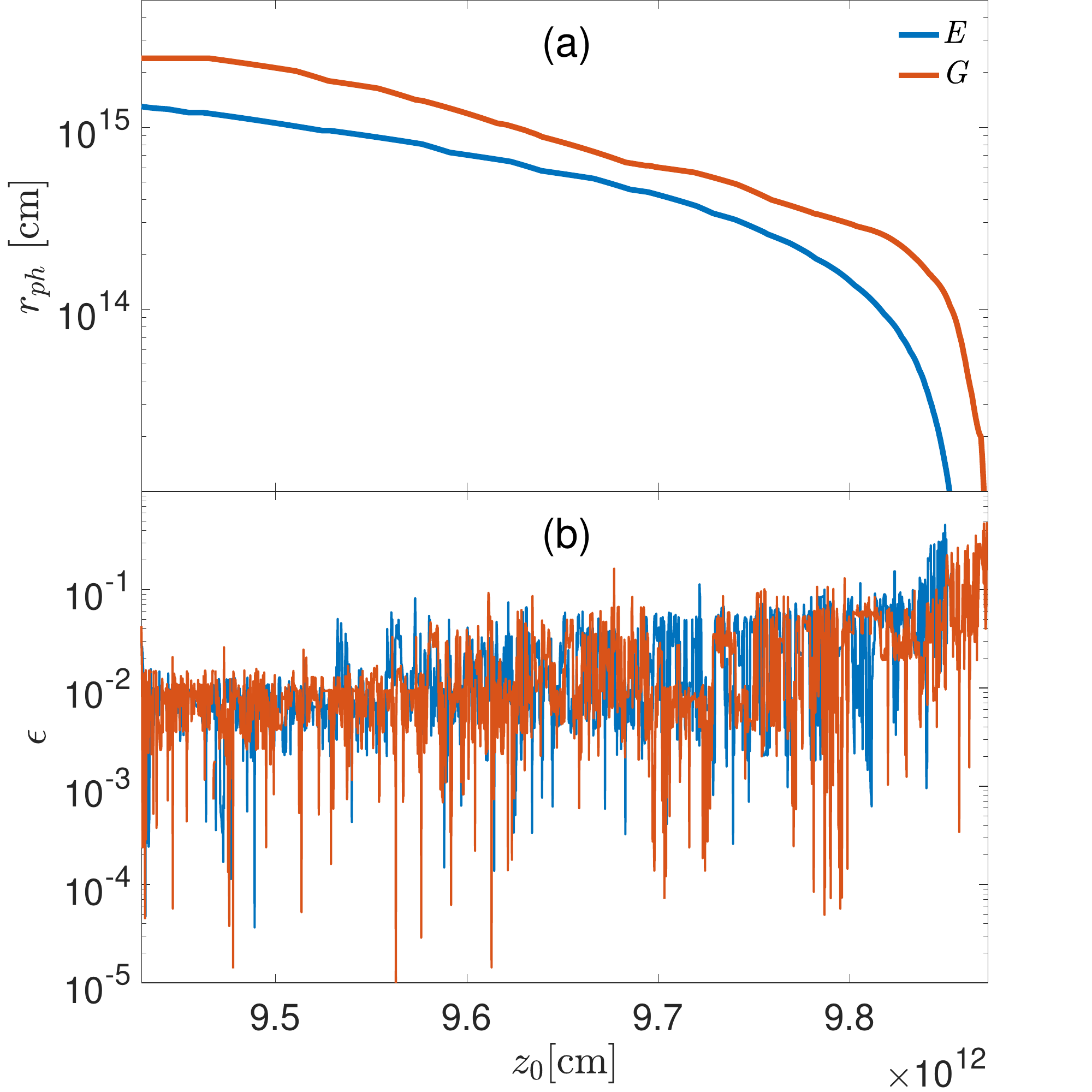}
		\includegraphics[scale=0.26]{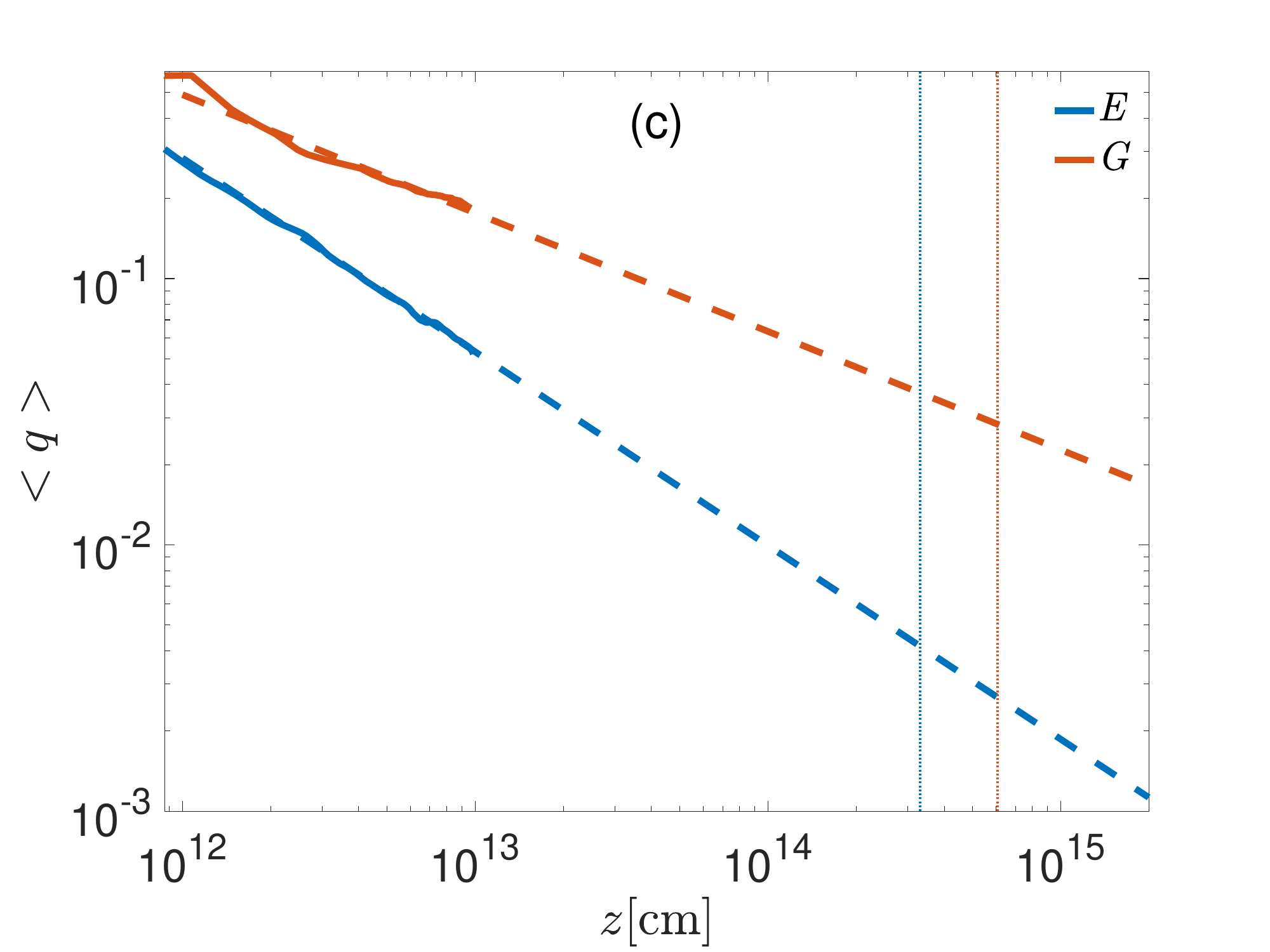}
		\caption[]{
			Photospheric radius (a) and radiative efficiency, $\epsilon=q_{ph}=(h_{ph}-1)/h_{ph}$ (b),
			of fluid elements along the jet axis in models $\E$ and $\G$, computed by extrapolating the data in the last snapshot of the 2D simulation.  The horizontal axis gives the distance from the origin of each fluid element at the end of the 2D simulation.  The shown slab 
			%			(a)+(b): The jet elements at the last snapshot of the 2D simulations of models $ \E $ and $ \G $, 
			(from $ \sim 9.4\times 10^{12}\cm $ to $ \sim 9.9\times 10^{12}\cm $) is the entire jet. 
			%Shown are the extrapolated photospheric radius (a) and the photospheric efficiency (b).
			(c): Energy-weighted mean value of $q=(h-1)/h$ plotted as a function of the 
			distance of the jet front from the origin. 
			The solid lines delineate the values obtained from the 2D simulation.  
			The dashed lines correspond to power-law extrapolation of the 2D data.
			The vertical dotted lines mark the energy-weighted mean photospheric radius in each model.
		}
		\label{fig:ph_eps}
	\end{figure}

	The excessive baryon contamination in the intermittent jets pushes the photosphere of the outflow to radii significantly
	larger than those computed in the continuous jet model ($\sim 10^{12}$ cm).  We find that for most fluid elements the optical depth at the end of the 2D simulation (at $r\simeq10^{13}$ cm) is still large, $\tau \sim 10^3$.  Using the extrapolation method developed in GLN19 beyond this radius yields $r_{ph}$. 
	%enhances the optical depth substantially, thereby pushing the photosphere of jet elements to large radii.
	Figure \ref{fig:ph_eps}a depicts the extrapolated photospheric radii of fluid elements on the jet axis for models $ \E $ and $ \G $, showing that the photosphere of most fluid elements is located very far out, at $ z > 10^{14}\cm $.  The horizontal axis gives the distance of jet elements in the last snapshot of the 2D simulation from the 
	injection point; the shown slab (from $ \sim 9.4\times 10^{12}\cm $ to $ \sim 9.9\times 10^{12}\cm $) 
	constitutes the entire jet.

	The photospheric radius thereby obtained is found to be much larger than 
	the coasting radius of the jet, $r_{coast}\lesssim 10^{12}$ cm.  This readily implies very low radiative efficiency, as indeed confirmed by direct calculations.
	Figure \ref{fig:ph_eps}b depicts the radiative efficiency of fluid elements on the jet axis in models $ \E $ and $ \G $ as a function of their location in the last 2D snapshot.  As seen, the radiative efficiency of 
	most fluid elements in both models is extremely low, with the exception of the section near the jet's head, where the efficiency approaches $ \epsilon \approx 10\% $. The overall efficiency depends also on the relative energy carried by each fluid element, which is not shown in this plot. In fact, we find that  
	the jet section where $\epsilon$ is highest contains only a small fraction ($ \sim 1\% $) of the total jet energy, hence its contribution to the net emission is small.  To estimate the overall efficiency we average $\epsilon$
	over the entire jet energy.  
	Figure \ref{fig:ph_eps}c shows the energy-weighted mean value of $q$ on the jet axis as a function of the jet's head location. The solid line in each model corresponds to the mean value of $q$ in the 2D simulation data.  The decline of $<q>$ with radius
	reflects gradual conversion of internal energy stored in shells of high specific enthalpy (or high $\eta$) 
	to bulk energy of the heavily loaded shells (with low $\eta$) through the decay of internal shocks.     
	As explained above, the 2D simulation box does not extend all the way to the photosphere.  Hence, in order to estimate the efficiency we extrapolated the $<q>$ profile obtained in the simulation (roughly a power law) to the energy-weighted mean photosphere (marked as the dotted vertical lines in the figure). The value of $<q>$ at $r_{ph}$ is the energy-weighted efficiency $<\epsilon>$. As seen, the overall efficiency is a fraction of a percent in model $\E$ and about $2\%$ in model $\G$.
	This should be contrasted with continuous jet models for which the overall efficiency was found to be high in GLN19.
	For the other intermittent jet models in table \ref{table} the overall efficiency is expected to be much lower than
	in models $\E$, $\G$.  
	
	The above analysis suggests that an intermittent engine that expels a hydrodynamic outflow cannot produce a detectable $\gamma$-ray signal if its duty cycle is significantly shorter than the jet breakout time from the stellar envelope.

	\subsection{Afterglow}
	\label{sec:photospheric}
	
	While intermittent hydrodynamic jets maintain a terminal Lorentz factor of a few dozens and will not produce a detectable $ \gamma $-ray emission, they may still yield a detectable afterglow signal at later times as they interact with the external medium.
	Previously, GRB jets with high baryon loading, known as ``dirty fireballs" \citep{Dermer2000,Huang2002,Rhoads2003,Lamb2017a}, have been suggested as the origin for an on-axis orphan afterglow \citep{Nakar2003}.
	
	Consider a relativistic jet with isotropic equivalent energy $ E_{\rm iso} $ and terminal Lorentz factor $ <\eta> $, which propagates in an external medium with a density profile $ \rho(r) = \rho_a (r/r_i)^{-\alpha} $. Our intermittent jet simulations show that $ <\eta> \lesssim 30 $, so that the reverse shock, which is formed from the interaction with the medium, is Newtonian \citep{Nakar2004}. In such case the fireball equations dictate that the jet starts decelerating at radius
	\begin{equation}\label{eq:r_dec}
	R_{\rm dec} = \bigg[\frac{E_{\rm iso}(3-\alpha)}{4\pi \rho_a c^2\eta^2r_i^\alpha}\bigg]^{\frac{1}{3-\alpha}}~,
	\end{equation}
	and the afterglow peak takes place at time
	\begin{equation}\label{eq:t_obs}\small
	t_{\rm obs} = \bigg[\frac{(3-\alpha)E_{\rm iso}\eta^{8-2\alpha}}{2^{5-\alpha}\pi\rho_a c^{5-\alpha}r_i^\alpha}\bigg]^{\frac{1}{3-\alpha}}(1+z)
	\equiv A \bigg(\frac{E_{\rm iso,52}\eta_{200}^{8-2\alpha}}{\rho_{a,0}}\bigg)^{\frac{1}{3-\alpha}}(1+z)~,
	\end{equation}
	where $ z $ is the redshift, $E_{iso}=E_{\rm iso,52} \cdot 10^{52}$ erg and $\eta=200\eta_{200}$.
	For a homogeneous medium ($ \alpha = 0 $) the normalization is set as $ A \sim 14\s $.
	%(see also \citealt{Ghirlanda2018} for a discussion of corrections for the value of $ A $).
	In such medium Equation (\ref{eq:t_obs}) indicates that the afterglow peak time is delayed for lower Lorentz factors as $ t_{\rm obs} \propto \eta^{-8/3} $. That implies that for a given energy and density, plugging in the terminal Lorentz factor obtained in our simulations with $ \lmin = 0 $ postpones the afterglow onset time to $ \sim $ day after the jet launch. For the terminal velocities found when $ \lmin > 0 $, the afterglow rises on $ \sim $ hour timescales.
	
	The rate of on-axis orphan afterglows can be constrained by observations. \citet{Nakar2003} have used X-ray fluxes to show that under the assumption that such dirty fireballs have similar energy per solid angle as regular GRBs, the rate of on-axis orphan afterglows cannot be much higher than that of typical GRB afterglows. Similar results were obtained by \citet{Ho2018} based on the lack of fast fading optical orphan afterglows in the iPTF survey.
	%\citet{Ghirlanda2018} have used optical observations of afterglows onset to show that $ \sim 20\% $ of GRB jets maintain a Lorentz factor of $ \Gamma < 100 $ at the afterglow onset.
	Orphan on-axis afterglows have possibly been observed before in the optical bands \citep[e.g.][]{Cenko2013}, although it may be that in this case we just missed the $\gamma$-rays from this event due to the partial $\gamma$-ray sky coverage (at any time less than half of the sky is covered by sensitive $\gamma$-ray detectors). It is challenging however to differentiate on-axis orphan afterglows from off-axis orphan afterglows which emerge for regular GRB jets seen off-axis.
	%multi-wavelength observations which show a delayed peak time in the radio bands with respect to other bands \citep{Lamb2017} or high radio variability \citep[see][and additional techniques therein]{Huang2002}.
	It may be possible, however, to distinguish a dirty fireball from an off-axis orphan afterglow by the different contribution of the reverse shock during the rise of the afterglow \citep{Sari1999a}.

	\section{Summary \& Discussion}
	\label{sec:discussion}
	
	The central engines of GRBs have long been considered to be intermittent, with duty cycles typically shorter than 
	the outflow breakout time from the envelope of the collapsed star. 
	The variability of the engine naturally leads to formation of strong internal shocks in the outflow that 
	under appropriate conditions can dissipate a considerable fraction of the bulk energy near or above the photosphere. 
	Such an activity has been commonly invoked to account for the nonthermal spectra and rapid variations of 
	the prompt GRB emission although, as shown recently (GLN19), fast variability can also arise from sporadic 
	mixing of jet  and cocoon material even in cases of steady jet injection. 
	
	Since the photosphere is located well above the star, it is imperative to understand how a modulated jet 
	evolves as it propagates through the envelope of the progenitor.  In this paper we addressed this question by 
	performing 3D simulations of hydrodynamic jets injected  periodically from a small radius inside the star, for  
	a range of modulation amplitudes and periods.  By further evolving the outflow from the outer edge of the 
	3D simulation box (at $5R_\star$) to a much larger radius ($100R_\star$) using 2D simulations, we were able 
	to compute the radiative efficiency of the flow. 
	
	We find, quite generally, that modulated jets with periods substantially shorter than the breakout time suffer excessive mass loading that strongly inhibits their emission.  For all the models investigated, 
	the radiative efficiency during the prompt phase never exceeded $\sim 1\%$, and in most cases is 
	much smaller.  This should be contrasted with emission from a continuous jet (steadily injected), that features
	highly efficient photospheric emission despite moderate mixing (GLN19). 
	The excessive loading in the periodic jets is caused by dragging of stellar material by shocks that form at the interfaces between high and low power shells.  Mixing by instabilities at the collimation zone, which is the primary loading mechanism in a continuous jet, is subdominant in the periodic jets. 
	
	Our results challenge the variable engine model for long GRBs. 
	The question whether similar behavior is expected also in short GRBs (sGRBs) remains open.
	The combination of shorter breakout times and less massive medium surrounding sGRB jets is expected to mitigate the propagation of the relativistic jet through the medium.
	Therefore, the modulations considered here, of timescales which are comparable to the breakout times of sGRBs, may still yield an efficient GRB emission. However, if shorter duty-cycles lead to heavier loading, as found in our simulations, modulations of $ T \lesssim 10^{-2}$s might provide similar results for sGRBs as well.
	
	Our main conclusion is that intermittent hydrodynamic outflows cannot produce the observed prompt $\gamma$-ray emission, but may appear as X-ray, optical and radio transients that are similar to GRB afterglows starting hour to days after the GRB. Mild magnetization may alter these results, but it is not clear at present how.  
	3D RMHD simulations \citep{Gottlieb2020b} show that sub-dominant magnetic fields can suppress hydrodynamic instabilities in a continuous jet \citep[see also][for a qualitative discussion]{Matsumoto2019}, thereby preventing significant mass loading of the jet . If this holds also in case of periodic jets, it could be that some critical magnetization is needed to produce a successful GRB.
	However, as we have shown the dominant loading process in periodic jets is not associated with hydrodynamic instabilities and, therefore, the results found for steady magnetized jets cannot	be directly applied to intermittent jets.  Attempts to perform 3D MHD simulations of periodic jets with mild magnetization are currently underway.

	\section*{Acknowledgements}
	This research was supported by the Israel Science Foundation (grant 1114/17). OG and EN were partially supported by an ERC grant (JetNS).	
	\bibliographystyle{mnras}
	\bibliography{Modulations}
	
	\appendix
	
	\section{Convergence}
	\label{sec:convergence}
	
\begin{figure}
		\centering
		\includegraphics[scale=0.26]{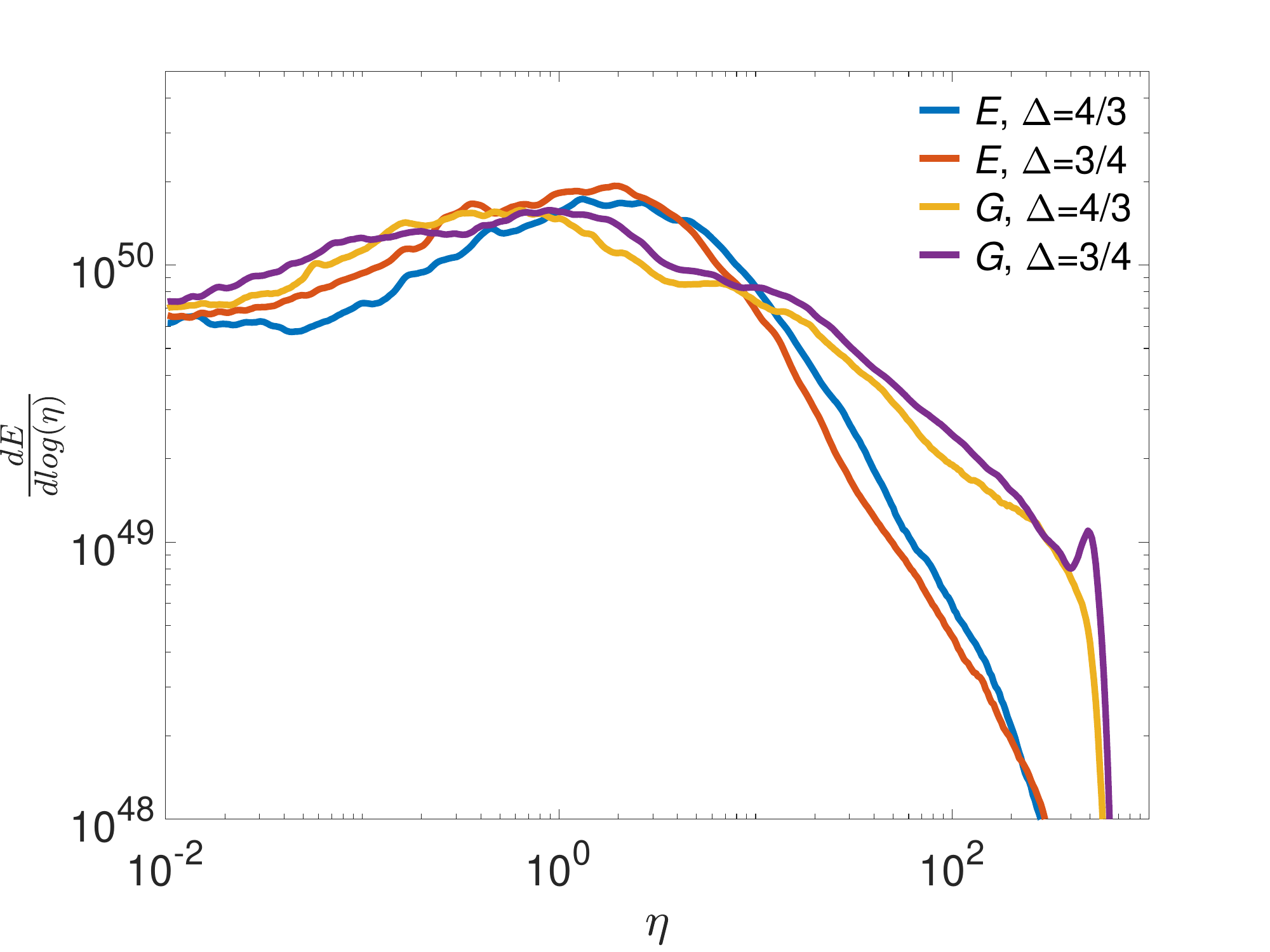}
		\caption[]{
			A comparison of distributions of the energy per logarithmic scale of $ \eta $ from simulations with different cells size $ \Delta $.
		}
		\label{fig:convergence}
	\end{figure}

	We carry out convergence tests for simulations $ \E $ and $ \G $ as these the models which hold promise for producing the brightest electromagnetic signals. We perform two additional simulations for each model. One with a lower grid resolution, in which the cells size is 4/3 larger than the original size. And one of a higher resolution, in which the cells size is 3/4 the size of the original cells size.
	We test the convergence by comparing the energy distribution per a logarithmic scale of $ \eta $, as depicted in Figure \ref{fig:convergence}.
	Although the cells size in the two convergence simulations differ from each other by almost a factor of two, the distributions agree well with differences up to $ \lesssim 30\% $. That implies that the results found in our simulations are qualitatively similar when increasing or decreasing the grid resolution.

	\label{lastpage}
\end{document}